\documentclass[useAMS,usenatbib]{mn2e}
\usepackage{graphicx}

\title[A Stellar Prominence in QS Vir]{A stellar prominence in the white dwarf/red dwarf binary QS Vir: evidence for a detached system}

\author[S. G. Parsons et al.]{S.~G.~Parsons$^{1}$\thanks{steven.parsons@warwick.ac.uk},
T.~R.~Marsh$^{1}$,
B.~T.~G{\"a}nsicke$^{1}$,
and C.~Tappert$^{2}$ \\
$^{1}$Department of Physics, University of Warwick, Coventry, CV4 7AL\\
$^{2}$Departamento de Astronom{\'i}a y Astrof{\'i}sica, Pontificia Universidad Cat{\'o}lica, Vicu{\~n}a Mackenna 4860, 782-0436 Macul, Chile}
\begin{document}
\input{pjw_aas_macros.cls}
\date{Accepted 2010 November 23.  Received 2010 November 23; in original form 2010 November 1}

\pagerange{\pageref{firstpage}--\pageref{lastpage}} \pubyear{2010}

\maketitle

\label{firstpage}

\begin{abstract}
Using high resolution UVES spectra of the eclipsing Post Common Envelope Binary QS Vir we detect material along the line of sight to the white dwarf at orbital phase $\phi=0.16$. We ascribe this to a stellar prominence originating from the M dwarf secondary star which passes in front of the white dwarf at this phase. This creates sharp absorption features in the hydrogen Balmer series and Ca {\sc ii} H and K lines. The small size of the white dwarf allows us to place tight constraints on the column density of hydrogen in the $n=2$ level of $\log_{10}{N_2} = 14.10 \pm 0.03\, \mathrm{cm}^{-2}$ and, assuming local thermodynamical equilibrium, the temperature of the prominence material of $\sim 9000$K. The prominence material is at least $1.5$ stellar radii from the surface of the M dwarf. The location of the prominence is consistent with emission features previously interpreted as evidence for Roche lobe overflow in the system. We also detect Mg {\sc ii} 4481{\AA} absorption from the white dwarf. The width of the Mg {\sc ii} line indicates that the white dwarf is not rapidly rotating, in contrast to previous work, hence our data indicate that QS Vir is a pre-cataclysmic binary, yet to initiate mass transfer, rather than a hibernating cataclysmic variable as has been suggested.

\end{abstract}


\section{Introduction}

A binary system initially composed of two main-sequence stars, in an orbital period of between $\sim$10-1000 days \citep{kolb96}, will result in the two stars orbiting within a common envelope of material when the more massive primary star evolves off the main-sequence and expands beyond its Roche lobe. The effect of this common envelope phase is to reduce the orbital period of the system dramatically. The resulting system, now composed of a tightly bound white dwarf and low-mass main-sequence star, evolves towards a semi-detached state via the loss of orbital angular momentum. Eventually, the surface of the main-sequence star will touch its Roche lobe and initiate mass transfer on to the white dwarf via a gas stream that passes through the L1 point, creating a cataclysmic variable (CV) system.

The evolution before and during the CV stage is driven by angular momentum loss via gravitational radiation (\citealt{kraft62}; \citealt{faulkner71}) and magnetic braking (\citealt{verbunt81}; \citealt{rappaport83}). However, whilst the angular momentum loss from gravitational radiation is well understood, angular momentum loss rates via magnetic braking are still uncertain and only recently has it become possible potentially to measure the angular momentum loss rate directly from post common envelope binaries (PCEBs) \citep{parsons10}. \citet{schreiber10} recently found evidence of disrupted magnetic braking in PCEBs.

Mass loss from nova eruptions during the CV stage are expected to result in a widening of the binary separation and an increase in the Roche radius of the donor star \citep{shara86}. This can result in the donor losing contact with its Roche lobe and thus mass transfer ceases and the systems becomes detached once more, until angular momentum once more brings the two stars close enough to re-initiate mass transfer. The existence of these hibernating CVs is controversial, however, if they exist they would resemble detached pre-cataclysmic binaries. 

QS Vir (EC $13471-1258$) is a close eclipsing binary system containing a DA white dwarf and a dMe dwarf. \citet{odonoghue03} presented photometric monitoring and time-resolved optical and ultra-violet spectroscopy of QS Vir and proposed that the system was a hibernating CV rather than a pre-CV. Their evidence for this was the large rotational velocity of the white dwarf ($V_\mathrm{rot,WD} = 400 \pm 100$ km$\,$s$^{-1}$), which implied that there had been an earlier phase of spin-up by mass transfer, and the behaviour of the multi-component H$\alpha$ emission-line profile, which they concluded showed evidence of a weak mass transfer steam. Recently \citet{ribeiro10} questioned this conclusion, arguing that the emission feature seen close to the white dwarf in the Doppler images of \citet{odonoghue03} is the result of wind accretion onto the white dwarf rather than a result of a low-level accretion stream shock. Their data also provided no indication that QS Vir was a hibernating CV. It is important to establish the evolutionary status of QS Vir since its eclipsing nature makes it ideal to study angular momentum loss.

Here we present high-resolution UVES spectra of QS Vir and use them to identify and characterise a stellar prominence. We then discuss the role of the prominence within the system and evolutionary status of QS Vir as a whole. 

\section{Observations and their reduction}

\begin{table}
 \centering
  \caption{Journal of VLT/UVES spectroscopic observations.}
  \label{spec_obs}
  \begin{tabular}{@{}lcccc@{}}
  \hline
  Date&Start&Exp     &Orbital&Conditions\\
      &(UT) &time (s)&phase  &(Transparency, seeing)\\
 \hline
 2001/08/19&00:07&300.0 &0.16&Good, $\sim 1$ arcsec\\
 2002/04/23&01:45&300.0 &0.05&Good, $\sim 1.5$ arcsec\\
 2007/08/17&00:46&1800.0&0.41&Variable, $\sim 2$ arcsec\\
\hline
\end{tabular}
\end{table}

Three spectra of QS Vir were obtained using the Ultraviolet and Visual Echelle Spectrograph (UVES) installed at the European Southern Observatory's Very Large Telescope (ESO VLT) 8.2-m telescope unit on Cerro Paranal in Chile \citep{dekker00} within both the 167.D-0407 (2001 and 2002) and 079.D-0276 (2007) programs, details of these observations are listed in Table~\ref{spec_obs}.

The reduction of the raw frames was conducted using the most recent standard pipeline release of the UVES Common Pipeline Library (CPL) recipes (version 5.0.1) within ESORex, the ESO Recipe Execution Tool, version 3.7.2. The standard recipes were used to optimally extract each spectrum. A ThAr arc lamp spectrum was used to wavelength calibrate the spectra, flux calibration was achieved using observations of standard stars. We then barycentrically corrected the wavelength scales of each of the spectra. The 2001 and 2002 spectra were obtained using UVES in a dichroic mode as part of the ESO Supernova Ia Progenitor Survey \citep{napiwotzki01}. They cover a wavelength range of $\sim 3500$ to $6650${\AA} with $\sim 80${\AA} gaps at 4580{\AA} and 5640{\AA}, the spectral resolution is around $R \sim 18500$. The 2007 spectra cover a different wavelength range of 3300-4520{\AA} in the blue arm and 5680-7515{\AA} and 7660-9440{\AA} in the red arm thus H$\alpha$ is not covered. These spectra have a resolution of $R \sim 70000$.

We determined the orbital phases of the 2001 and 2002 spectra using the ephemeris of \citet{odonoghue03}. \citet{parsons10} showed that the eclipse times of QS Vir vary substantially after 2002 hence for the 2007 spectra we used the following linear ephemeris fitted to the eclipse times after 2006:
\[\mathrm{MJD(BTDB)}= 48689.135\,529(54) + 0.150\,757\,6116(14) E.\]

\section{results}
\subsection{Spectral features}

\begin{figure*}
  \begin{center}
    \includegraphics[width=0.95\textwidth]{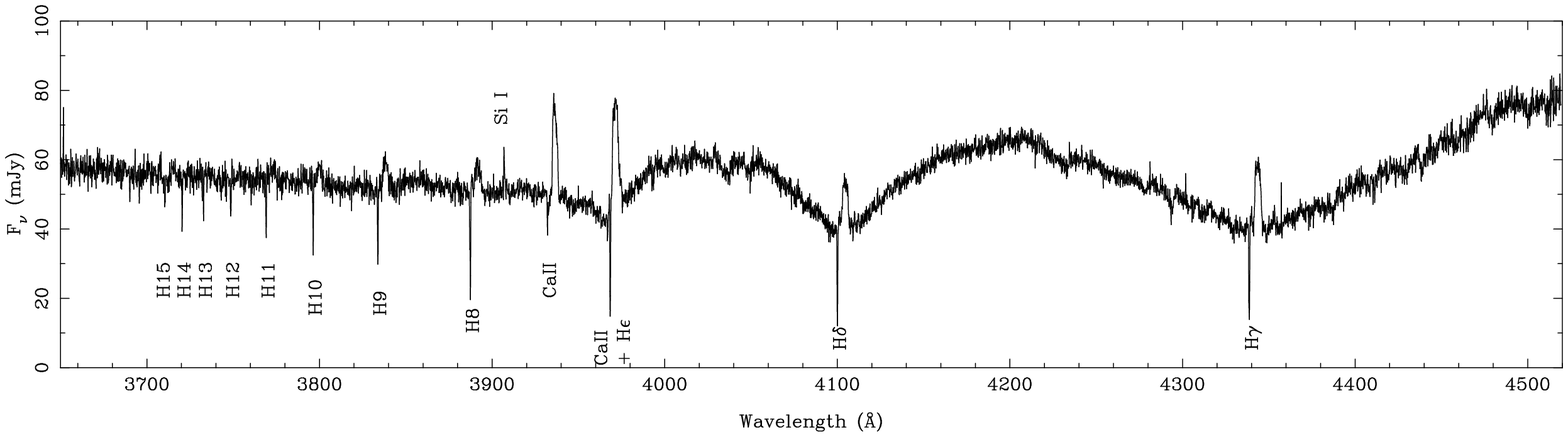}
    \includegraphics[width=0.95\textwidth]{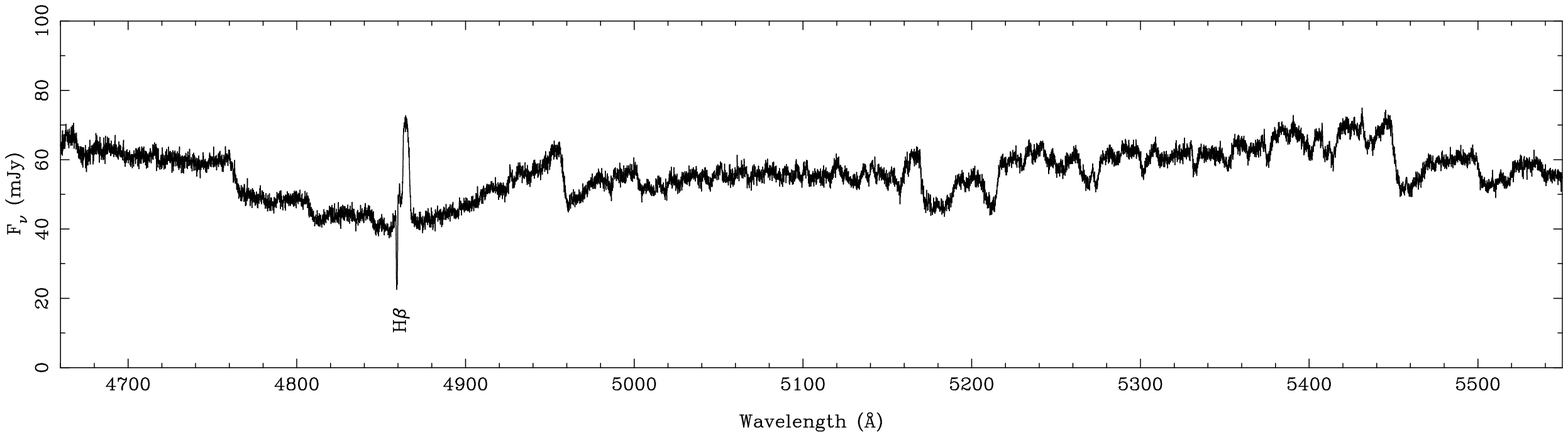}
    \includegraphics[width=0.95\textwidth]{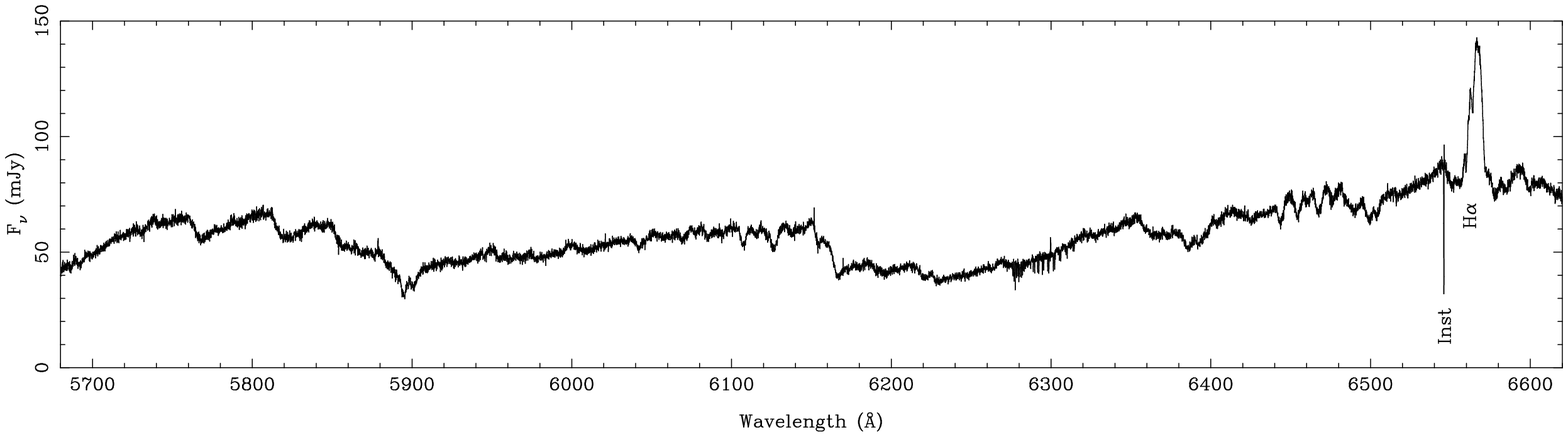} \\
  \caption{2001 UVES spectrum of QS Vir with the prominence features labelled. The sharp dip in the bottom panel labelled ``inst'' is an instrumental feature. Hydrogen Balmer, Ca {\sc ii} and Si {\sc i} emission from the red star is also visible (the silicon feature is labelled) as well as many absorption features from the red star. There is also weak Mg {\sc ii} 4481{\AA} absorption though this is not visible on this scale.}
  \label{specs}
  \end{center}
\end{figure*}

Figure~\ref{specs} shows the 2001 spectra. At short wavelengths, where the white dwarf dominates, broad Balmer absorption from the white dwarf is present. Additionally, hydrogen Balmer, Ca {\sc ii} H and K and Si {\sc i} 3906{\AA} emission from the red star is evident. The Si {\sc i} emission is irradiation driven therefore it appears quite narrow in the 2001 spectrum since the heated face of the secondary star is not fully visible at the orbital phase that this spectrum was taken. At long wavelengths absorption features from the red star dominate the spectrum. Similar features are seen in the 2002 and 2007 spectra. 

The 2001 spectra also contain sharp absorption features corresponding to the hydrogen Balmer series and Ca {\sc ii} H and K lines; these are labelled in Figure~\ref{specs}. These sharp absorption features are not present in the 2002 or 2007 spectra. We also detect Mg {\sc ii} 4481{\AA} absorption in all the spectra (see section~\ref{discus} for a discussion of this feature).

\subsection{The origin of the sharp absorption features}
\label{origin}

The sharp absorption features, labelled in Figure~\ref{specs}, are seen only in the 2001 spectra which means that they are either a transient feature or only visible at specific orbital phases. The depth of these features indicate that they cannot originate from the photosphere of the red star because they are deeper than the total contribution from the red star, therefore it is light originating from the white dwarf which is being absorbed. Hence this absorption is from material cooler than the white dwarf, between us and the white dwarf at an orbital phase of 0.16.

If this absorption is caused by a disk of material orbiting the white dwarf then the small change in direction of the material as it crosses the face of the white dwarf (slightly blueshifted on one side and slightly redshifted on the other) will broaden the absorption lines. Hence the width of the narrowest line, $\Delta V$, can be used to estimate the minimum distance of the material from the white dwarf, $R$, we find
\begin{eqnarray}
\Delta V \sim \frac{K_\mathrm{sec}}{\sqrt{1+q}}\left(\frac{a R_\mathrm{WD}^2}{R^3}\right)^{0.5},
\end{eqnarray}
where $K_\mathrm{sec}$ is the projected orbital velocity of the M dwarf (266 km$\,$s$^{-1}$), $q = M_\mathrm{sec} / M_\mathrm{WD}$, the mass ratio (0.52), $a$ is the orbital separation ($1.28 R_{\sun}$) and $R_\mathrm{WD} \sim 0.011 R_{\sun}$ is the radius of the white dwarf (all values from \citealt{odonoghue03}). The width of the narrowest line is 18 km$\,$s$^{-1}$ (see Section~\ref{prom_rvs}) hence the distance of the material from the white dwarf is at least $R > 0.22a$. Given the relatively low inclination of QS Vir ($\sim 75^\circ$, \citealt{odonoghue03}) it is unlikely that these absorption features are the result of disk material since it would have to be located far from the white dwarf and far out of the orbital plane.

The velocities of the absorption features are consistent with the white dwarf's radial velocity (see Section~\ref{prom_rvs} for radial velocity measurements). This implies that the material is locked in rotation with the binary since any material rotating within the system, along the line of sight of the white dwarf, will have the same radial velocity as the white dwarf. 

based on these arguments the most likely origin of this absorption is material within a stellar prominence originating at the red star. Stellar prominences have been seen in a number of different systems; slingshot prominences have been observed in the cataclysmic variables BV Cen \citep{watson07}, IP Peg and SS Cyg \citep{steeghs96} and long-lived loop prominences have been detected on the secondary star in AM Her \citep{gansicke98}. \citet{dunstone06} found a prominence from the K3 dwarf Speedy Mic (BO Mic) which shows similar features to those seen here and \citet{peterson10} discovered a large coronal loop in the Algol system. Since the red star in QS Vir is known to be active \citep{odonoghue03}, prominences originating at this star are certainly possible. These are the first observations of prominence material in absorption within a binary; all of the previous examples show sharp emission features from prominence material within binaries. This provides useful diagnostics of the physical conditions within the prominence.

\section{Modelling of the prominence}
The ratio of the equivalent widths (EWs) of the various absorption lines arising from the prominence can be used to constrain its temperature and density. Since the red star in QS Vir is close to filling its Roche lobe \citep{odonoghue03} these kind of features are important for understanding any accretion processes taking place in the system. 

\begin{figure*}
  \begin{center}
    \includegraphics[width=0.95\textwidth]{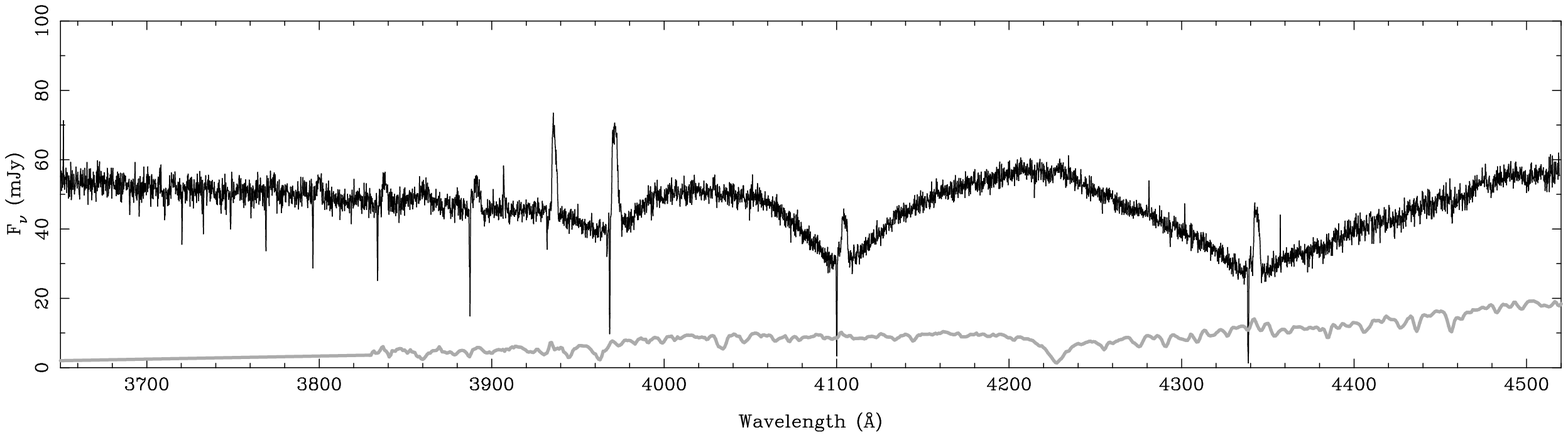}
    \includegraphics[width=0.95\textwidth]{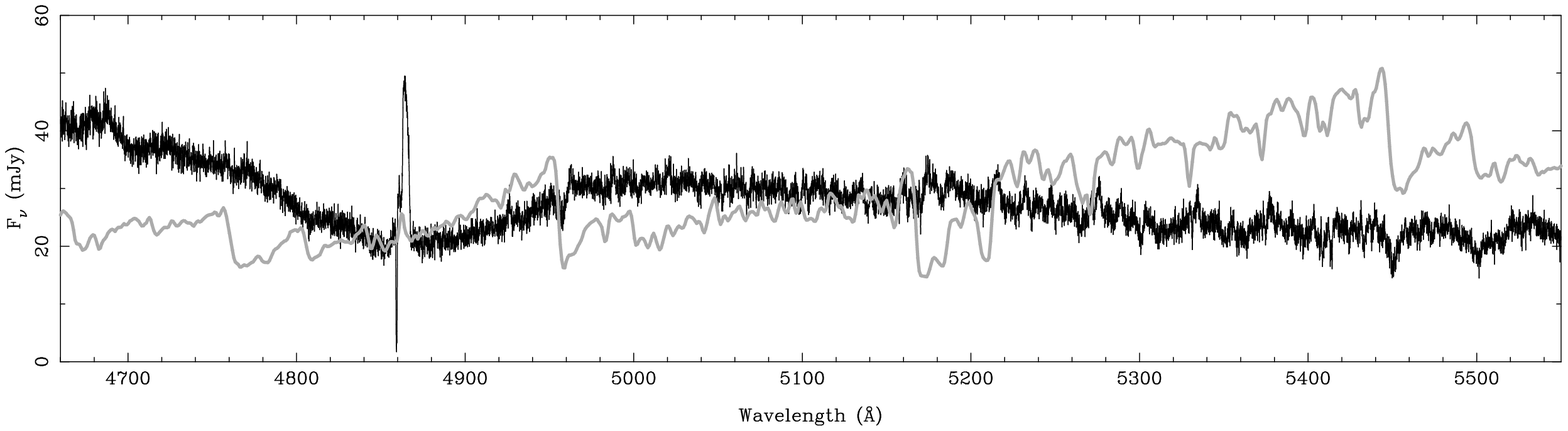}
    \includegraphics[width=0.95\textwidth]{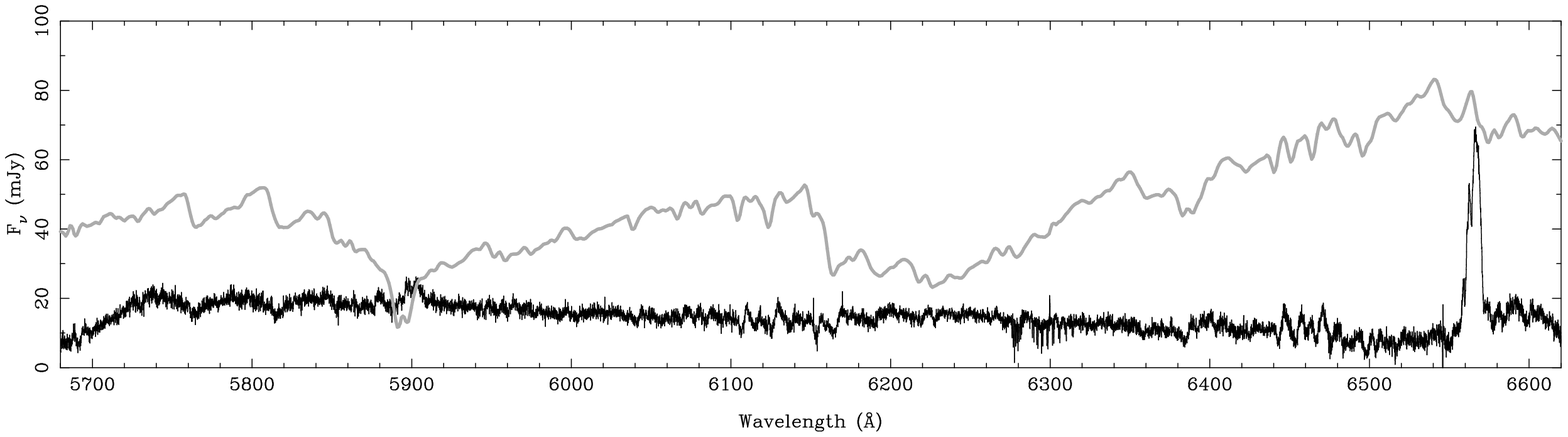} \\
  \caption{The 2001 spectrum with the red star's contribution subtracted. The \emph{grey} line is the M3 template that was subtracted from the original spectrum. This model starts at a wavelength of 3830{\AA} hence before this point we extrapolate the continuum level, since the white dwarf dominates at these wavelengths this has a minimal effect. The red star features have not been completely removed since the resolution of the template is less than the UVES spectra resolution but the overall flux level from the red star has been removed. The emission features were not included in the model and therefore have not been removed either. The small contribution from the white dwarf around the H$\alpha$ absorption makes determining an accurate equivalent width difficult.}
  \label{rd_sub}
  \end{center}
\end{figure*}

As we remarked earlier the depth of the prominence absorption in the higher Balmer series is far deeper than the total light emitted by the red star at these shorter wavelengths hence the prominence material must be blocking light from the white dwarf. However, in order to determine accurate EWs, the contribution from the red star has to be removed. We determined the red star's contribution to the flux across the entire wavelength range covered by using a model spectrum of a 14,000K $\log{g} = 8.25$ white dwarf (calculated using TLUSTY and SYNSPEC; \citealt{hubeny95}) and a range of M dwarf template spectra \citep{rebassa07} from spectral type M3 to M4 in order to closely match the parameters found by \citet{odonoghue03}. We used these to derive synthetic fluxes for both of the stars and adjusted the overall level to match the relative magnitudes shown in \citet{parsons10} for the ULTRACAM $u'$, $g'$, $r'$ and $i'$ filters. We fitted the ratio of the in-eclipse and out-of-eclipse flux levels for the $g'$ and $r'$ bands since these best cover the Balmer series. There was little difference between the spectral types used for the secondary, however, the M3 model combined with the white dwarf model best fitted all of the filter ratios. Figure~\ref{rd_sub} shows the 2001 spectra with the M3 model, corrected to the radial velocity of the red star, subtracted off. Also shown is the M3 model which gives an indication of the relative contributions of the two stars at different wavelengths.

For the H$\delta$ to H$\alpha$ lines, the absorption from the prominence occurs on top of the emission from the secondary star because the emission becomes wider in these lines. Additionally, the H$\epsilon$ absorption lies on top of the Ca {\sc ii} emission from the red star. Ideally, in order to correct the EWs for this emission, we would subtract off the emission using observations of the lines at a similar phase without the absorption component. However, since the UVES spectra cover only three different orbital phases, none of which are close to each other, we cannot subtract the emission from the spectra accurately, and therefore we use the white dwarf model to determine the continuum level and correct the EWs accordingly. This approach works well for the H$\epsilon$ to H$\beta$ lines where the absorption occurs in the wings of the emission, but unfortunately the H$\alpha$ absorption lies right on the strong emission peak and an accurate correction would require knowledge of the emission profile. Since this is unknown, we correct the absorption to the white dwarf continuum but note that there is likely to be an additional correction required hence all measurements of this line will be unreliable. Additionally, as Figure~\ref{rd_sub} shows, the contribution from the white dwarf around H$\alpha$ is much lower than the red star's contribution making the absorption feature weaker still.

\subsection{EWs of the absorption lines}
\label{prom_rvs}
We measured the EW, velocity offset and full-width-half-maximum (FWHM) of all of the prominence features. We determined the velocity offset and FWHM by fitting the line with a combination of a Gaussian and a straight line. The results of this are listed in Table~\ref{ews}

After correcting for the red star emission, we find that, within the uncertainties, the fluxes in the centres of the sharp absorptions of H$\delta$, H$\gamma$ and H$\beta$ are zero thus the prominence must be large enough to block our view of the entire white dwarf, and also the material in the prominence cannot be very hot. A weighted average of the velocities of the lines gives the radial velocity of the absorption as $-123.3 \pm 0.2$ km$\,$s$^{-1}$. The radial velocity of the white dwarf for this phase is $-118 \pm 11$ km$\,$s$^{-1}$ (\citealt{odonoghue03}, systemic velocity included). As previously mentioned, this is evidence of the absorption being from a prominence since any material in solid body rotation within the system, along the line of sight of the white dwarf, must have the same radial velocity as the white dwarf.

\begin{table}
 \centering
  \caption{Measured parameters of the prominence absorption lines. ``C Fac'' is the additional equivalent width needed to correct for the emission from the secondary star.}
  \label{ews}
  \begin{tabular}{@{}lcccc@{}}
  \hline
  Line ID&Velocity      &FWHM         &EW    &C Fac \\
         &km$\,$s$^{-1}$&km$\,$s$^{-1}$&(m{\AA})&(m{\AA})\\
 \hline
 H$\alpha$ 6562.760   & $-114.1 \pm 1.3$ & $74.0 \pm 4.2$ & $389 \pm 30$ & 600 \\
 H$\beta$ 4861.327    & $-119.4 \pm 0.6$ & $40.5 \pm 1.5$ & $604 \pm 47$ & 160 \\
 H$\gamma$ 4340.465   & $-125.0 \pm 0.4$ & $35.5 \pm 1.1$ & $506 \pm 28$ & 60  \\
 H$\delta$ 4101.735   & $-120.9 \pm 0.3$ & $31.9 \pm 0.9$ & $455 \pm 32$ & 15  \\
 H$\epsilon$ 3970.074 & $-126.9 \pm 0.4$ & $38.2 \pm 1.3$ & $400 \pm 29$ & 5   \\
 Ca {\sc ii} 3968.469 & $-120.0 \pm 2.0$ & $28.6 \pm 5.3$ & $89 \pm 26$  & 0   \\
 Ca {\sc ii} 3933.663 & $-121.7 \pm 0.9$ & $18.0 \pm 2.3$ & $82 \pm 20$  & 0   \\
 H8 3889.055          & $-128.6 \pm 0.6$ & $30.1 \pm 1.5$ & $322 \pm 28$ & 0   \\
 H9 3835.397          & $-121.4 \pm 0.7$ & $28.6 \pm 1.8$ & $232 \pm 25$ & 0   \\
 H10 3797.910         & $-123.0 \pm 0.8$ & $25.9 \pm 1.8$ & $167 \pm 25$ & 0   \\
 H11 3770.634         & $-121.0 \pm 1.0$ & $28.2 \pm 2.4$ & $159 \pm 22$ & 0   \\
 H12 3750.152         & $-130.8 \pm 1.8$ & $29.3 \pm 4.4$ & $125 \pm 23$ & 0   \\
 H13 3734.372         & $-124.9 \pm 1.4$ & $25.7 \pm 3.5$ & $89 \pm 19$  & 0   \\
 H14 3721.948         & $-127.9 \pm 1.5$ & $28.1 \pm 3.6$ & $98 \pm 17$  & 0   \\
 H15 3711.971         & $-118.6 \pm 2.6$ & $33.8 \pm 6.5$ & $86 \pm 19$  & 0   \\

\hline
\end{tabular}
\end{table}

\subsection{Column densities}
In this section we describe the model we used to determine the column densities of hydrogen and calcium in the prominence and constrain its temperature and turbulent velocity.

\subsubsection{The model}
We calculate the line absorption profile using a Voigt profile of the form:
\begin{eqnarray}
\label{line_prof}
\alpha_\lambda = \frac{e^2\lambda_0^2f}{m_ec^2} \left( \frac{1}{\Delta\lambda_D\sqrt{\pi}}e^{-(\Delta\lambda / \Delta\lambda_D)^2/2} * \frac{1}{\pi}\frac{\gamma}{\Delta\lambda^2 + \gamma^2} \right),
\end{eqnarray}
where $e$ is the electron charge, $\lambda_0$ is the central wavelength of the line, $f$ is the oscillator strength, $m_e$ is the mass of an electron, $c$ is the speed of light, the star symbol is the convolution operator and $\Delta\lambda = \lambda - \lambda_0$, the distance from the centre of the line. The first term inside the brackets is a Gaussian profile representing the thermal and microturbulent broadening, in this case
\begin{eqnarray}
\label{dlamd}
\Delta\lambda_D = \frac{\lambda_0}{c}\left( \frac{kT}{m} + \xi^2\right)^{0.5},
\end{eqnarray}
where $k$ is the Boltzmann constant, $T$ is the temperature, $m$ is the mass of the atomic species involved and $\xi$ is the microturbulence. This is convolved with the Lorentzian profile on the right in Eqn~\ref{line_prof} which represents the natural line broadening where $\gamma$ is the FWHM given by
\begin{eqnarray}
\label{gamma}
\gamma = \frac{\lambda_0^2}{4\pi c}\sum\limits_{j<i}A_{ij},
\end{eqnarray}
where the $A_{ij}$ are the Einstein A-coefficients contributing to the line. This gives us the absorption profile per atom.

The optical depth is then given by multiplying this by the column density ($N$ - the total column density of the atomic species in a given energy level):
\[\tau_\lambda = \alpha_\lambda N.\]
Hence integrating over the line profile gives us the EW via
\begin{eqnarray}
\label{ewidth}
\mathrm{EW} = \int (1 - e^{-\tau_\lambda})d\lambda.
\end{eqnarray}
Therefore the only variables are the temperature, $T$, the turbulent velocity, $\xi$, and the column density, $N$. However, there is a degeneracy between the thermal and turbulent broadening whereby $V_\mathrm{Turb}^2 + V_\mathrm{Therm}^2 = C$, a constant. For large background sources the EW is also dependent upon the covering fraction (the amount of the emission source covered by the absorber), which leads to a degeneracy between the column density and the covering fraction. Fortunately, since the white dwarf is small it is completely covered (as can also be seen in the very deep absorption cores in Figure~\ref{rd_sub}) hence the degeneracy is broken and we are able to obtain reliable column densities.

\subsubsection{Column density of hydrogen in the $n=2$ level}

\begin{figure}
  \begin{center}
    \includegraphics[width=0.95\columnwidth]{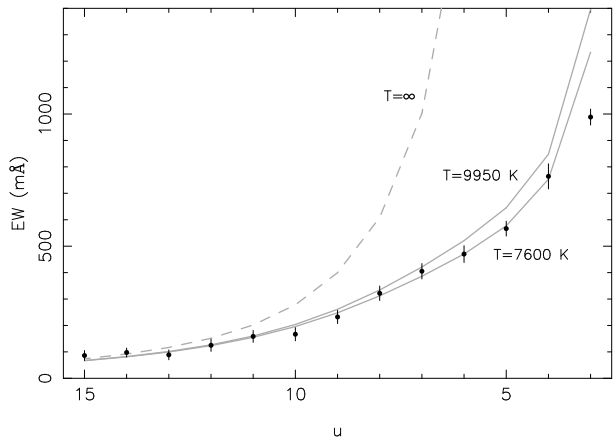} \\
  \caption{Model fits to the EWs of the Balmer series. ``$u$'' is the upper energy level involved in the transition (H$\alpha =3$, H$\beta =4$ etc). For temperatures below 7600K the EWs can be fitted by increasing the turbulence. The 9950K model shows the three sigma upper limit to fitting the EWs, any additional increase in temperature causes the lower lines of the series to become unsaturated. Hence a good fit is still obtained if there is no turbulence and a high temperature (as is favoured by ionisation equilibrium considerations). Due to the uncertainty in the correction required to the H$\alpha$ line due to emission from the red star we do not fit this line. The dashed line is $\int \tau_\lambda d\lambda$ which shows the predicted EW without the effects of saturation; the higher lines of the series show very little saturation.}
  \label{ew_fit}
  \end{center}
\end{figure}

We fitted our model to the measured EWs of the hydrogen Balmer absorption lines (with the exception of H$\alpha$) using Levenberg-Marquardt minimisation \citep{press86} for a range of turbulent velocities. We find that $N_2 = 1.26 \pm 0.10 \times 10^{14}$cm$^{-2}$.  For temperatures below 7600K the microturbulence can balance the thermal broadening thus offering no constraint. The fit to the EWs for a temperature of 7600K is shown in Figure~\ref{ew_fit}. By using a range of turbulent velocities we also find that $V_\mathrm{Turb}^2 + V_\mathrm{Therm}^2 = 63.2$ (km$\,$s$^{-1})^2$. The three sigma upper limit on the temperature from fitting the EWs is 9950K, as also shown in Figure~\ref{ew_fit}. Above this temperature the lower lines of the series become unsaturated and the EWs are overestimated. However, these high temperatures (above 7600K) require zero turbulence to fit. 

\subsubsection{Column density of Ca {\sc ii}}
Since we also detect the calcium H and K lines from the prominence we can carry out the same analysis as for the hydrogen Balmer lines in order to establish the column density of Ca {\sc ii}. Unfortunately, since we have only two lines with rather large uncertainties in the EWs we cannot constrain the column density as tightly as with the Balmer lines.

We calculated the range of column densities that will fit the observed EWs of the Ca {\sc ii} lines (to within $2\sigma$) over a grid of temperatures from 3000--11,000K and turbulent velocities from 0--8 km$\,$s$^{-1}$. We find that high turbulent velocities ($>4$ km$\,$s$^{-1}$) favour low column densities ($11.3 < \log_{10}{N_\mathrm{CaII}} (\mathrm{cm}^{-2}) < 11.6$) for all temperatures. However, lower turbulence leads to a much larger spread in column densities ($11.5 < \log_{10}{N_\mathrm{CaII}} (\mathrm{cm}^{-2}) < 14.6$) with a much larger dependence on temperature whereby the lower the temperature, the higher the column density. 

\begin{figure}
  \begin{center}
    \includegraphics[width=0.95\columnwidth]{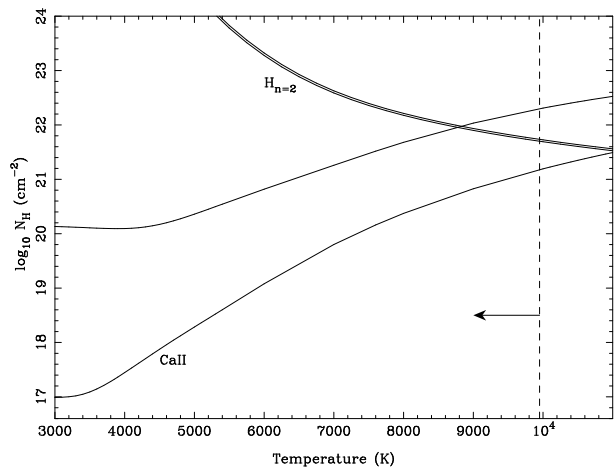} \\
  \caption{Column density of hydrogen in the prominence as a function of temperature. Using the Ca {\sc ii} lines permits a large range of column densities over the temperature range. Using the Balmer lines places tighter constraints on the column density. The dashed line shows the three sigma upper temperature constraint from fitting the Balmer lines. The permitted column density of hydrogen from the Balmer lines and Ca {\sc ii} lines overlap at a higher temperature ($\sim 8800$K) which is still permitted from fitting the Balmer lines but requires very low turbulence.}
  \label{column}
  \end{center}
\end{figure}

\subsubsection{Column density of hydrogen in the $n=1$ level}
\label{colden}
We can constrain the column density of neutral hydrogen using our constraints on the column densities of hydrogen in the $n=2$ state and calcium. We assume a size of the prominence of $10^{10}$cm (see section~\ref{psize}) and solar abundances. From the Saha-Boltzmann equation (assuming local thermodynamical equilibrium) we can determine the column density of neutral hydrogen that would give the observed column density of calcium. We can also use the column density of hydrogen in the $n=2$ state to calculate the column density of neutral hydrogen via a Boltzmann factor. The results of both of these calculations are shown in Figure~\ref{column} over a range of temperatures. The large uncertainty in the column density of calcium results in a large range of possible neutral hydrogen column densities. However, the hydrogen in the $n=2$ state gives a much tighter constraint. The two regions overlap at temperatures above 8800K implying that the prominence material is hot. This temperature is below the three sigma limit set by the EWs of the Balmer lines and implies low levels of turbulence (as seen in Figure~\ref{ew_fit}). It is also possible that the material may not be entirely in local thermodynamical equilibrium. However, our various constraints are consistent with the prominence having little or no turbulence and a temperature of $\sim 9000$K. These results are consistent with observations of other stellar prominences \citep{dunstone06} and modelling of solar prominences (\citealt{gilbert06}; \citealt{labrosse01}).

\begin{figure*}
  \begin{center}
    \includegraphics[width=0.4\textwidth]{pos1.eps}
    \includegraphics[width=0.4\textwidth]{pos2.eps} \\
  \caption{\emph{Left:} a visualisation of the system as viewed from above. The dotted lines show the Roche lobes of the two stars and the path that a stream would take were one present. Also marked are the positions of the L4 and L5 points, the two straight lines show the viewing angle of the white dwarf from Earth at the start and end of the exposure, thus the prominence material is located within the shaded area. \emph{Right:} a velocity map of QS Vir with the same features as the left panel. The star shows the location of the emission feature found by \citet{odonoghue03}, the circle represents the uncertainty in its position.}
  \label{visualise}
  \end{center}
\end{figure*}

\section{Discussion}

\subsection{The size of the prominence}
\label{psize}
We can place a lower limit on the size of the prominence since we know that the white dwarf is completely covered hence the prominence must be at least as large as the white dwarf. However, we can do better than this since we know that the white dwarf is obscured for the full 300 second exposure therefore the minimum size of the prominence must be $2 \pi \Delta \phi R = 0.145 R$, where $\Delta \phi$ is the change in phase during the exposure and $R$ is the distance of the prominence from the white dwarf. The width of the narrowest feature tells us that $R > 0.22a$ (see section \ref{origin}) hence the size of the prominence must be at least $0.22a \times 0.145 = 0.032a \sim 3.7R_\mathrm{WD}$. It could of course be considerably larger than this limit.

A minimum distance of the prominence from the red star can be calculated since we know that the prominence lies along the line of sight from the white dwarf at an orbital phase of 0.16. This gives a minimum distance of the prominence from the centre of the red star of $0.81a$. \citet{odonoghue03} estimate $R_\mathrm{RD} = 0.42 R_{\sun}$ which gives a minimum distance of the prominence above the surface of the red star as $0.62 R_{\sun}$, almost $1.5$ stellar radii away.

Limits can be placed on the density of the material within the prominence. From section~\ref{colden} we determined a range of column densities of neutral hydrogen of $0.5 \times 10^{22} < N_H (\mathrm{cm}^{-2}) < 1.0 \times 10^{22}$. Using our minimum prominence size (0.032a) gives an upper limit on the density of $n_H = 3.5 \times 10^{12} \mathrm{cm}^{-3}$. A loose lower limit can be calculated by assuming that the prominence is smaller than the binary separation, this gives a density of $n_H =5.6 \times 10^{10} \mathrm{cm}^{-3}$. This range of densities is consistent with observations  and modelling of both solar \citep{ballegooijen10} and other stellar prominences \citep{schroeder83}.

The fact that QS Vir is an eclipsing system and we know that the prominence passes in front of the white dwarf means that it must also pass in front of the red star. There is no evidence in the 2002 or 2007 spectra of any prominence features; this could either be due to the prominence being a short-lived feature or that we have not covered the phase at which the prominence crosses the face of the red star. Observations at a range of phases including the detection of the orbital phase at which the prominence passes in front of the red star would tightly constrain the size, location and mass of the prominence. 

\subsection{Evidence for a detached system}
\label{discus}

Figure~\ref{visualise} shows a visualisation of the system. All the features were drawn using the best estimate system parameters from \citet{odonoghue03}, we used a mass ratio of $q=0.52$ and $K_\mathrm{WD}+K_\mathrm{sec}=137+266=403$ km$\,$s$^{-1}$. We note here that this is not an exact visualisation, due to the uncertainty in the system parameters, but it serves to understand better the location of the prominence and other features previously identified in this system.

The left hand panel of Figure~\ref{visualise} shows the system in spatial co-ordinates, we also show the positions of the Roche lobes of the two stars and the route that a stream would take (were one present). The shaded region indicates the potential positions of the prominence (it must lie along the line of sight of the white dwarf at $\phi=0.16$). Interestingly, it is possible for the prominence to be located close to the L5 point. Since this is an unstable, though neutral point, the material would still have to be supported in some way (e.g. by a magnetic field). Similar features have been identified by \citet{jensen86} in the eclipsing PCEB V471 Tau. They found dips in the soft X-ray flux from material along the line of sight at orbital phases consistent with the material being at both the L4 and L5 points. They attribute this material to very large coronal loops anchored at the K star. \citet{wheatley98} also found X-ray dips in V471 Tau but caused by material much closer to the K star (within a stellar radius).

The right hand panel of Figure~\ref{visualise} shows the system in velocity co-ordinates. The same features as the left hand panel are shown (Roche lobes and stream) as well as the possible positions of the prominence (assuming that it is locked to the red star and hence the binary). We also indicate the position of an emission feature seen in the Doppler image of \citet{odonoghue03} at $(V_x,V_y)=(75,-75)$ km$\,$s$^{-1}$ the circle around it is the uncertainty in its position given the resolution of their Doppler image. This feature was interpreted by \citet{odonoghue03} as emission from an accretion stream, however a stream would not be located at these co-ordinates in velocity space, rather it would be located outside the Roche lobe of the white dwarf along the nearly diagonal line to the upper-right of the primary's Roche lobe as shown in Figure~\ref{visualise}. Therefore the emission feature detected by \citet{odonoghue03} cannot be from an accretion stream. 

The range of possible positions for the prominence passes through the emission feature, so, rather than being due to a low level accretion stream shock, the emission feature seen in the Doppler image could well be related to a prominence. If the prominence we have detected is due to the same material as seen in the Doppler map then it is remarkably long lived since the spectra used to create the Doppler map in \citet{odonoghue03} were taken between April 1992 and June 1993 almost a decade before the UVES spectra presented here, however, further data are needed to test this.

These observations support the argument made by \citet{ribeiro10} that the material physically inside the white dwarf's Roche lobe is likely to be the result of wind accretion onto the white dwarf. Their claim that this may be facilitated by a prominence-like magnetic loop appears to be supported by our data. Our observations, combined with those of \citet{ribeiro10}, show that QS Vir is currently a detached system.

\begin{figure}
  \begin{center}
    \includegraphics[width=0.95\columnwidth]{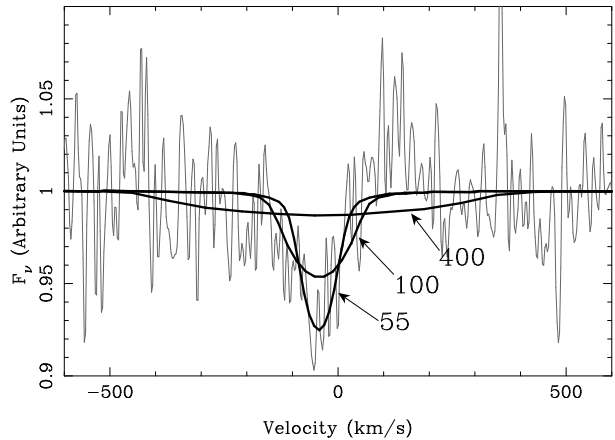} \\
  \caption{Mg {\sc ii} 4481{\AA} absorption from the white dwarf seen in 2007 in slightly poor conditions. Overplotted are three models with varying $v\sin{i}$ values of $55$ km$\,$s$^{-1}$, $100$ km$\,$s$^{-1}$ and $400$ km$\,$s$^{-1}$. The $v\sin{i} = 55$ km$\,$s$^{-1}$ model fits the line best but the quality of the data is not good enough to tightly constrain its value. However, the narrow nature of this feature confirms that the white dwarf is not a rapid rotator.}
  \label{mag_ab}
  \end{center}
\end{figure}

Mg {\sc ii} 4481{\AA} absorption is present in all of the spectra, though it is quite weak in the 2001 and 2002 spectra. The strength of the absorption in all three spectra are consistent within their uncertainties. The feature is best seen in the 2007 data due to the longer exposure time and is shown in Figure~\ref{mag_ab}. Since this absorption is seen in all the spectra it cannot have the same origin as the sharp features seen only in the 2001 spectra. Furthermore, the velocity of the line is in the opposite sense to the red star meaning that this absorption comes from the atmosphere of the white dwarf. This feature potentially allows us to measure the radial velocity of the white dwarf, however, with just three spectra a measurement of the radial velocity is not possible with the data to hand. QS Vir's current evolutionary stage (pre-CV or hibernating CV) can also be addressed by our spectra; the identification of Mg {\sc ii} 4481{\AA} absorption from the white dwarf allows us to constrain the rotational velocity of the white dwarf. 

\citet{odonoghue03} determined a rotational velocity for the white dwarf of $V_\mathrm{rot} = 400 \pm 100$ km$\,$s$^{-1}$ from modelling of {\it Hubble Space Telescope}/STIS ({\it HST}/STIS) UV spectra, suggesting that the white dwarf had been spun-up by an earlier phase of mass transfer. However, this measurement is marginal since it is at the limit of the {\it HST}/STIS spectral resolution. Figure~\ref{mag_ab} shows the Mg {\sc ii} 4481{\AA} absorption feature in the 2007 spectrum with three models with different $v\sin{i}$ values for the white dwarf overplotted (calculated using TLUSTY; \citealt{hubeny95}). The model with $v\sin{i} = 400$ km$\,$s$^{-1}$ overestimates the width of the line and even with a rotational velocity of $100$ km$\,$s$^{-1}$ the predicted line width is too wide, a rotational velocity of $v\sin{i} = 55$ km$\,$s$^{-1}$ fits the line best. However, the white dwarf moves by $\sim 100$ km$\,$s$^{-1}$ during the 1800 second exposure which can entirely explain the width of the line, meaning that a $v\sin{i} = 0$ is favoured. Nevertheless, we can rule out a large $v\sin{i}$ value, which was one of the main arguments of \citet{odonoghue03} that QS Vir is a hibernating CV. Hence, the fact that there is no evidence of mass transfer in the system, leads us to conclude that QS Vir has yet to reach a semi-detached configuration and is therefore a pre-CV rather than a hibernating CV.

\section{Conclusions}

The main sequence star in the eclipsing PCEB QS Vir is very close to filling its Roche lobe. \citet{schreiber03} showed that QS Vir is one of the few PCEBs that has passed through most of its PCEB life time, this makes it a particularly interesting system since the majority of PCEBs are young systems that have only just emerged from their common envelope phase. 

\citet{odonoghue03} found an emission feature from within the Roche lobe of the white dwarf which, combined with their measurements of a rapidly rotating white dwarf, led them to believe that QS Vir was in fact a hibernating CV that had become detached as a result of a nova explosion. Here we have used UVES spectroscopy to identify and characterise a stellar prominence in the system which lies along the line of sight of the emission feature seen in \citet{odonoghue03} leading us to conclude that the emission feature may be associated with a prominence rather than a low-level accretion stream. It is also possible that the prominence may be located at the L5 point of the system. We have also detected Mg {\sc ii} absorption arising from the atmosphere of the white dwarf; the width of this feature is incompatible with rapid rotation of the white dwarf suggesting that QS Vir has yet to become a CV and is thus a pre-CV.

\section*{Acknowledgements}
TRM and BTG acknowledge support from the Science and Technology Facilities Council (STFC) grant number ST/F002599/1. The results presented in this paper are based on observations collected at the European Southern Observatory under programme IDs 167.D-0407 and 079.D-0276. We used the National Aeronautics and Space Administration (NASA) Astrophysics Data System. This research has made use of the National Institute of Standards and Technology (NIST) Atomic Spectra Database (version 3.1.5).

\bibliographystyle{mn2e}
\bibliography{qsvir}

\label{lastpage}

\end{document}